\documentstyle[12pt]{article}

\begin{document}

\pagestyle{empty}
\pagestyle{plain}
\setcounter{page}{1}

\def \a{\alpha}
\def \b{\beta}
\def \eps{\epsilon}
\def \d{{\rm d}}
\def \e{{\rm e}}
\def \beq{\begin{equation}}
\def \eeq{\end{equation}}

\def \ln{{\rm ln}}
\def \ov{\over}
\def \t{\tau}
\def \ol{\overline}
\def \alfa{\alpha}
\def \b{\beta}
\def \r{\right}
\def \s{\sigma}
\def \l{\left}
\def \la{\langle}
\def \ra{\rangle}
\def \eps{\epsilon}
\def \ab2{\alpha\beta^2}
\def \Tr{\rm Tr}
\begin{titlepage}

\title{Saddle Points Stability in the Replica Approach Off
Equilibrium}
\author{ M.E. Ferrero and M.A. Virasoro}
\maketitle
\begin{center}
{Dipartimento di Fisica, Universit\`a di Roma,
  {\em La Sapienza},\\
I-00185 Roma, Italy\\
and\\
INFN Sezione di Roma I, Roma, Italy.}
\end{center}

\vspace{1truecm}

\begin{abstract}
We study the replica free energy surface for a spin glass model near the
glassy temperature. In this model the simplicity of the equilibrium solution
hides non trivial metastable saddle points.
By means of the stability analysis performed for one and two real replicas
constrained, an interpretation for some of them is achieved.
\end{abstract}

\end{titlepage}

\vspace{3truecm}

\section{Introduction}

In spin glass models, replica symmetry breaking implies the
existence of many pure states of equilibrium. Their hierarchical organisation
is described by  the order parameter $Q_{ab}$  \cite{MPV}.
In the SK model $Q_{ab}$ has a continuum form and it is a marginally
stable saddle point of the replica free energy surface in the
limit $ n \rightarrow 0$ \cite{ddk1}.
In other models, like the p-spin spherical model, the stable solution is just
one-step, yielding a simpler equilibrium  state organisation \cite{cris}.

In this paper we investigate this different feature from the stability point
of view. We use the generalisation of the replica method recently developed to
study non equilibrium states \cite{fpv} \cite{kpv}.
The general idea is to consider $R$ identical copies of the
model (real replicas) constrained to have specific mutual overlaps.
Then, by forcing the overlaps out of equilibrium,
one can probe the phase-space structure of
the original model and obtain information on some of its non equilibrium
states.

By performing the stability analysis in the replica approach off equilibrium,
we  uncover the physical  meaning for some saddle points other than
the one yielding the Gibbs-measure results.

\section{The unconstrained model and its saddle points}

\subsection{The model}

Let us consider the following truncated replica free energy density

\begin{equation}
F(Q)=-\frac{1}{n} (t\  Tr Q^{2}+\frac{1}{3} Tr Q^{3}+\frac{1}{6}
\sum_{a \neq b} q_{ab}^{3}+\frac{y}{12} \sum_{a \neq
b} q_{ab}^{4})
\label{F(Q)}
\end{equation}
where we  consider $t\sim T_c-T$ to be small.
The cubic term $\sum q^{3}_{ab}$, absent in the SK-model, gives
a  $y$ independent replica symmetry breaking.
A term of this kind appears in the
p-spin spherical model and in the Potts model \cite{gks}.
The $y$ parameter will play the role of a control parameter. We will assume
it to be small ($y<<t$) but positive or negative (in the SK model it is
positive).

In   \cite{gks} it was shown that for $y<0$ the
stable solution is simply a matrix with one replica symmetry breaking (1RSB).
The probability distribution of the overlap between equilibrium configurations
is then

\begin{equation}
P_{eq}(q)=m_{1} \delta (q)+(1-m_{1}) \delta (q-q_{EA})
\end{equation}
where $q_{EA}=2 t+\frac{10}{3} y t^2$  and $m_1= \frac{1}{2}+y
t$ are obtained
by {\it stationary conditions}.

In the solution we will consistently truncate the $O(y^2)$ terms. As a
consequence of the stationary conditions, the $O(y^2)$ terms in the solution
give $O(y^4)$ contribution to free energy.
In the following we will then consider free energy up to $O(y^3)$.

\subsection{Unstable saddle points}

For positive $y$ we know that a better approximation to the full solution is
given by the 2RSB saddle point

\begin{eqnarray}
q(x)&=& \left\{ \begin{array}{lll}
q_{0}=&0& \hbox{$0<x<m_1$} \\
q_{1}=&{t}+\frac{23}{12} y t^{2}& \hbox{$m_1<x<m_2$}\\
q_{2}=&{2t}+\frac{23}{6} y t^{2}&  \hbox{$m_2<x<1$}\\
                \end{array}
             \right. \nonumber\\
m_{1}&=&\frac{1}{2}+\frac{1}{2} y t \nonumber\\
m_{2}&=&\frac{1}{2}+\frac{3}{2} y t
\label{sol2rsb}
\end{eqnarray}
and free energy $F_{2RSB}>F_{1RSB}$ ($F_{2RSB}-F_{1RSB}\sim
y^2$).

To study fluctuations around this 2RSB S.P. we consider the eigenvalue
equation

\begin{equation}
(2 t+ q_{ab}+y q_{ab}^{2}+\lambda) f_{ab}+\{Q,f\}_{ab} =0
\label{eq1rr}
\end{equation}
where $f=Q-Q_{SP}$. This equation, analysed in details in the context of the
SK model  \cite{ddk2}, can be solved through the explicit construction of
the eigenvectors exploiting the symmetry of the equation.
In the notation of \cite{ddk2} the eigenvectors are classified as longitudinal
 ($f_{ab}$), anomalous ($f^{\mu}_{ab}$) and replicon ($f^{\mu \nu}_{ab}$).
For positive $y$ the minimum eigenvalue is $\lambda=-\frac{yt^2}{6}$ and it is
well known that unstable directions belong also to the replicon family.

Let us discuss the analytic continuation to negative $y$ of this
solution, that we expect to have new unstable directions. It leads to
$m_2<m_1$, a result which does not allow the usual
probabilistic interpretation.

We found the minimum eigenvalue to be negative ($\lambda=\frac{5yt^2}{6}$).
The unstable eigenvectors belong to the
longitudinal and anomalous families. They are

\begin{eqnarray}
longitudinal\rightarrow f_{ab}&=& \left\{ \begin{array}{ll}
f_{0}(0)=0& \hbox{$if\;\;\;\; a\cap b=0$} \\
f_{0}(1)\neq0& \hbox{$if\;\;\;\; a\cap b=1$}\\
f_{0}(2)\neq0&  \hbox{$if\;\;\;\; a\cap b=2$}\\
                \end{array}
             \right. \nonumber\\
anomalous\rightarrow f_{ab}^{\mu}&=& \left\{
\begin{array}{ll}
f_{1}(0)=0& \hbox{$if\;\;\; a\cap b=0$} \\
f_{1}(1)\neq 0& \hbox{$if\;\;\; a\cap b=1
\;\;\;max(a\cap\mu,b\cap\mu)=0$} \\
f_{1}(2)\neq 0& \hbox{$if\;\;\; a\cap b=2
\;\;\;max(a\cap\mu,b\cap\mu)=0$}\\
f_{1}(1)(1-\frac{n}{m_1})&  \hbox{$if\;\;\; a\cap b=1 \;\;\;
max(a\cap\mu,b\cap\mu)>0$}\\
f_{1}(2)(1-\frac{n}{m_1})&  \hbox{$if\;\;\; a\cap b=2 \;\;\;
max(a\cap\mu,b\cap\mu)>0$}\\
                \end{array}
             \right. \nonumber\\
\end{eqnarray}
where, as in \cite{ddk2}, $a\cap b=k$ if $q_{ab}=q_k$.

For future reference, we anticipate that by considering additional RSB
the new unstable S.P. will have $m_{i+1}-m_{i}\sim y$.
There is, of course, the analytic continuation (to negative $y$) of
the continuum solution (slope $=\frac{1}{y}$), which is the only one
marginally stable for positive $y$. It has obviously
a free energy greater than the other solutions.

\section{The constrained model}

Let us consider $R$ Real Replicas and introduce in the partition function
$\frac{R}{2}(R-1)$ complex Lagrangean multipliers
($\epsilon_{rs}$) to keep the mutual overlaps constrained ($q_c^{rs}$).
The effective averaged free energy has two contributions \cite{fpv}

\begin{equation}
F_{R}({\bf Q},q_c^{rs})=F({\bf Q})+F_{constr}(
\epsilon_a^{rs},q_c^{rs})
\end{equation}

The functional $F(\bf Q)$ has the same form  as for a single real
replica with ${\bf Q}$ a $Rn \times Rn$ matrix

\beq
 {\bf Q}=
 \bordermatrix{
     &   &   &   &    \cr
     &Q^{11} &P^{12} &... &P^{1R} \cr
     &P^{12^T} &Q^{22} &... &P^{2R}\cr
     &... &...&... &...\cr
     &P^{1R^T}&P^{2R^T}&...&Q^{RR}\cr
     }
\eeq
where $P^{rs}_{aa}=\epsilon_{rs}^{aa}$.
The "interaction" term $F_{constr}(\epsilon_a^{rs},q_c^{rs})$ reads

\beq
F_{constr}(\epsilon_a^{rs},q_c^{rs})=-
\frac{\beta^2}{2}\sum(q_c^{rs}-
\epsilon_a^{rs})^2
\label{F_{constr}}
\eeq

The stationary conditions are then

\beq
\frac{\partial F}{\partial q^{rr}_{ab}}=0 \;\;\;
\frac{\partial F}{\partial p^{rs}_{ab}}=0 \;\;\;
\frac{\partial F}{\partial p^{rs}_{aa}}=\beta^2 (p_{aa}^{rs}-
q^{rs}_c)
\label{stationary}
\eeq

In the following we consider the particular case $R=2$ and
we make for $Q$ and $P$ a Parisi ansatz.
Before analysing the solution, let us explain how we studied
the stability.

\subsection{2RR Stability}

To study  the Hessian of $F_2$ we consider $F$ at fixed $p_{aa}=p_d$.
{}From (\ref{stationary}) we see that the saddle point for $p_{aa}$,
which originally is integrated along the imaginary axis, lies then on the
real axis.
The integration path can be deformed to go through the saddle  point but
the integration path will be perpendicular to the real axis.
This S.P. has to a maximum when compared to values along the
integration path. Notice that this distinguish $p_d$ from all other matrix
terms which are integrated along the real axis.
The stability with respect to $p_d$ fluctuations around
the saddle point can be then verified in (\ref{F_{constr}}) and, being
not proportional to $t$, decouples.

The equation for $F_2$ is then

\begin{equation}
(2 t+{\bf q}_{\alpha\beta}+y {\bf q}_{\alpha\beta}^{2}+\lambda)
{\bf f}_{\alpha\beta
}+\{{\bf Q},{\bf f}\}_{\alpha\beta} =0
\label{eq2rr}
\end{equation}
where $\alpha\beta=1,...,2n$ but $\alpha-\beta\neq n$. In order to reduce this
problem to an ultrametric one we can observe that the symmetry group of the
Hamiltonian without constraint is $S_{2n}$ (all replicas are equivalent).
After imposing the constraint it becomes
$S_{n}\otimes (S_{2})^{\otimes n}$ (to permute
simultaneasly replicas in both systems or to permute equivalent replicas
between systems). Since all solutions found are symmetrical under the action
of $S_2$ on all the replicas,
we divide the eigenvectors in two superfamilies, the symmetrical and the
antisymmetrical one.
We then parametrise the fluctuation matrix
${\bf f}$ around the ${\bf Q}_{S.P.}$ in terms of

\begin{equation}
{\bf f}=\left(
\begin{array}{c c c}
+f^1& &+f^x\\
\\
+f^x& &+f^1
\end{array}
\right)
+
\left(
\begin{array}{c c c}
+f^z& &+f^y\\
\\
-f^y& &-f^z
\end{array}
\right)
\end{equation}
where the $f^{1,x,z}$ are $n*n$ symmetrical matrices  and
$f^y$ antisymmetrical.

The eigenvalue equations are now decoupled and read, for
$a\neq b$,

\begin{eqnarray}
(2 t+ q_{ab}+y q_{ab}^{2}+\lambda)
f^1_{ab}+\{Q,f^1\}_{ab}+\{f^x,P\}_{ab} & =&
0 \nonumber\\
(2 t+ p_{ab}+y p_{ab}^{2}+\lambda)
f^x_{ab}+\{P,f^1\}_{ab}+\{f^x,Q\}_{ab}&=&
0
\label{eq2rrsimm}
\end{eqnarray}

and

\begin{eqnarray}
 (2 t+ q_{ab}+y q_{ab}^{2}+\lambda)
f^z_{ab}+\{Q,f^z\}_{ab}+[f^y,P]_{ab} & =&
0 \nonumber\\
(2 t+  p_{ab}+y p_{ab}^{2}+\lambda)
f^y_{ab}+\{Q,f^y\}_{ab}+[f^z,P]_{ab}
 & = & 0
\label{eq2rrasimm}
\end{eqnarray}

The eigenvectors of (\ref{eq2rrsimm}) and (\ref{eq2rrasimm})
are then
respectively of the form

\begin{equation}
{\bf f}_{}=\left(
\begin{array}{c c c}
+f^1_{}& &+f^x_{}\\
\\
+f^x_{}& &+f^1_{}
\end{array}
\right)
\;\;
{\bf f}_{\mu}=\left(
\begin{array}{c c c}
+f^1_{\mu}& &+f^x_{\mu}\\
\\
+f^x_{\mu}& &+f^1_{\mu}
\end{array}
\right)
\;\;
{\bf f}_{\mu\nu}=\left(
\begin{array}{c c c}
+f^1_{\mu\nu}& &+f^x_{\mu\nu}\\
\\
+f^x_{\mu\nu}& &+f^1_{\mu\nu}
\end{array}
\right)
\end{equation}

\begin{equation}
{\bf f}_{}=\left(
\begin{array}{c c c}
+f^z_{}& &0\\
\\
0& &-f^z_{}
\end{array}
\right)
\;\;
{\bf f}_{\mu}=\left(
\begin{array}{c c c}
+f^z_{\mu}& &+f^y_{\mu}\\
\\
-f^y_{\mu}& &-f^z_{\mu}
\end{array}
\right)
\;\;
{\bf f}_{\mu\nu}=\left(
\begin{array}{c c c}
+f^z_{\mu\nu}& &+f^y_{\mu\nu}\\
\\
-f^y_{\mu\nu}& &-f^z_{\mu\nu}
\end{array}
\right)
\end{equation}
where $f^y_{\mu}$ and
$f^y_{\nu \mu}$ are antisymmetrical matrices breaking the
hierarchical subgroups of permutation group (as the 1RR ones).

\subsection{2RR constrained saddle points}

We investigate the possible 2RSB solutions in the range $0<p_d<q_{EA}$ $p_d=
p^{12}_{aa}$. We try to follow the solution by continuity when $p_d$ varies.
At different points in the interval a stable solution becomes unstable
and we have to search for the new stable solution.

Let us consider $y<0$.
We find four critical $p_d$ values $0<p^A<p^B<p^C<p^D<q_{EA}$.
At each critical $p_d$ value,  two different solutions coincide and zero modes
appear to modify their stability.

We start at $p_d \simeq 0$. The stable solution is

\begin{eqnarray}
q(x)&=& \left\{ \begin{array}{ll}
               0 & \hbox{$0<x<m_2$}  \\
           q_{EA}=2t+\frac{10}{3}yt^2 & \hbox{$m_2<x<1$}
           \end{array}
            \right. \nonumber\\
p(x)&=& \left\{ \begin{array}{ll}
              0  & \hbox{ $\ \ 0<x<m_2$} \\
             p_{d}  & \hbox{ $\ \ m_2<x<1$}
           \end{array}
            \right. \nonumber\\
m_2&= & \frac{1}{2}+y t
\label{0A}
\end{eqnarray}

and the free energy density is

\begin{equation}
F^{0A}-F^{unconstr}_{1RSB}=\Delta F^{0A}=\frac{1}{12}(-2 y t^2
p_{d}^2-p_{d}^3)+O(y^4)
\end{equation}

If $p_{d}=0$ the free energy is the same as the equilibrium one and
$\frac{\partial F^{0A}}{\partial p_d}=\frac{dF^{0A}}{dp_d}=0$. For small
$p_d$ the free energy is an increasing function of $p_d$  (fig.\ref{fig:1}).
The minimum eigenvalue is $\lambda=-\frac{2}{3}yt^2-p_d$.
Thus when $p_d>p^A=-\frac{2}{3}yt^2$ this solution becomes unstable and the
new stable solution is:

\begin{eqnarray}
q(x)&=&\left\{ \begin{array}{lll}
        q_{0}=&0 &\hbox{$0<x<m_1$}  \\
        q_{1}=&p_d-\frac{2}{3} y t (p_{d}-t)&\hbox{$m_1<x<m_2$}
\\
        q_{2}=&2 t+\frac{10}{3} y t^2-\frac{1}{6} y p_d (p_{d}-4t)&
\hbox{$m_2<x<1$}
        \end{array}
        \right. \nonumber\\
 p(x)&=&\left\{ \begin{array}{lll}
        p_{0}=&0   &\hbox{$\ \ \ \ \ \ \ \ \ \ \ \ \ 0<x<m_1$} \\
        p_{1}= &p_d-\frac{2}{3} y t (p_{d}-t)&\hbox{$\ \ \ \ \ \ \ \
\ \ \ \ \ m_1<x<m_2$} \\
        p_{2}=&p_{d}&\hbox{$\ \ \ \ \ \ \ \ \ \ \ \ \ m_2<x<1$}
        \end{array}
        \right. \nonumber\\
m_1&=&\frac{1}{4}+y\frac{p_{d}}{4}  \nonumber\\
m_2&=&\frac{1}{2}+y (t+\frac{p_{d}}{2})
\label{AB}
\end{eqnarray}

The free energy is

\begin{equation}
\Delta F^{AB}=\frac{y^2}{18}(2 p_d t^4-2 p_d^2 t^3+p_d^3 t^2-
\frac{1}{4}p_d^4
t)+O(y^3)
\end{equation}

This solution crosses, at $p_d=p^A$, the previous one.
For $p_d\sim -y$ the minimum eigenvalue is
$\lambda=\frac{p_d}{2}+\frac{yt^2}{3}$. As can be seen in  (fig.\ref{fig:1})
and (fig.\ref{fig:2}), $\Delta F^{AB}$ is an increasing function until
 $p_d=p_{max}=t+\frac{23}{12} y t^2$, where
$\frac{dF^B}{dp_d}=\frac{\partial F^{AB}}{\partial p_d}=0$.
For future reference we rewrite (\ref{AB}) on this specific value

\begin{eqnarray}
q(x)&=& \left\{ \begin{array}{lll}
       q_{0}=&0 &\hbox{$0<x<m_1$} \\
       q_{1}=&t+\frac{23}{12}y t^2&\hbox{$m_1<x<m_2$} \\
       q_{2}=&2 t+\frac{23}{6} y t^2 &\hbox{$m_2<x<1$}
        \end{array}
       \right. \nonumber\\
 p(x)&=& \left\{ \begin{array}{lll}
       p_{0}=&0   &\hbox{$\ \ \ \ 0<x<m_1$}\\
       p_{1}=&t+\frac{23}{12} y t^2&\hbox{$\ \ \ \ m_1<x<1$}
        \end{array}
       \right. \nonumber\\
m_1&=&\frac{1}{4}+y\frac{t}{4} \nonumber\\
m_2&=&\frac{1}{2}+y \frac{3 t}{2}
\label{Bmax}
\end{eqnarray}

The minimum eigenvalue is $\lambda=-\frac{yt^2}{6}$.
As $p_d$ increases further, $\Delta F^{AB}$ begins to decrease while all
eigenvalues read positive until the second critical value $p^B$.
The minimum eigenvalue for $p_d\simeq p^B$ is
$\lambda=-y(p_d^2-8 p_d t+8 t^2)/6$ and it corresponds to
fluctuations of $q_1$ and $p_1$. $p^B$ is defined by $\lambda|_{p^B}=0$.

For $p_{d}$ greater than $p^B$ the stable solution becomes

\begin{eqnarray}
q(x)&=& \left\{ \begin{array}{lll}
               q_{0}=&0 &\hbox{$0<x<m_1$}  \\
              q_{1}=&p_d-\frac{1}{9} y (-p_{d}^2+14 p_{d} t-14
t^2)&\hbox{$m_1<x<m_2$}\\
              q_{2}=&2 t+y \frac{10}{3} t^2 -\frac{p_d y}{6}(p_d-4 t)
&\hbox{$m_2<x<1$}
        \end{array}
       \right.\nonumber\\
 p(x)&=& \left\{ \begin{array}{lll}
         p_{0}=&0   &\hbox{$\ \ \ \ 0<x<m_1$} \\
        p_{1}=&p_d-\frac{1}{9} y  (p_{d}^2-2 p_{d} t+2
t^2)&\hbox{$\
\ \ \ m_1<x<m_2$} \\
        p_{2}= &p_{d}&\hbox{$\ \ \ \ m_2<x<1$}
     \end{array}
           \right. \nonumber\\
m_1&=&\frac{1}{4}+y \frac{p_d}{4}  \nonumber\\
m_2&=&\frac{1}{2}+y (t+\frac{p_{d}}{2})
\label{BC}
\end{eqnarray}

The free energy of this solution ($\Delta F^{BC}$), which differs from
(\ref{AB})  by terms of $O(y^3)$, is decrasing in $p_d$ (fig\ref{fig:2}).
Again at $p_d=p^B$ the two solutions crosses each other.
The minimum eigenvalue is $\lambda=y(p_d^2-8 p_d t+8 t^2)/6$. We have
$p_1-q_1=-2 y(p_d^2-8 pd t+8 t^2)/9\sim \lambda$ and the free
energy difference between (\ref{AB}) and (\ref{BC}), expressed in terms of
(\ref{BC}), reads $\sim(p1-q1)^3 \sim -y^3$.

We found (\ref{BC}) to be stable for $p^B<p_d<p^C \simeq 3/2 t$. In the range
$p^C<p_{d}<p^D< q_{EA}$ we did'nt find any 2RSB stable solution.
In particular (\ref{BC}) is unstable with respect to replicon fluctuations.
We expect that additional RSB lead to the stable solution.

Thus, let us consider $p_d>p^D=2t+\frac{23yt^2}{6}$. The stable solution is

\begin{eqnarray}
q(x)&=&\left\{\begin{array}{ll}
      0 &\hbox{$0<x<m_1$}  \\
     \frac{2}{3}(p_{d}+t)+y \frac{10}{81}
(p_{d}+t)^2&\hbox{$m_1<x<1$}
    \end{array}
      \right. \nonumber\\
 p(x)&=&q(x) \nonumber\\
m_1&=&\frac{1}{4}+y \frac{1}{6} (p_{d}+t)
\label{Dq}
\end{eqnarray}

The free energy, expressed in term of $\delta = p_d-q_{EA} \sim y$, reads
(fig.\ref{fig:3})

\begin{equation}
\Delta F^{Dq_{EA}}=\frac{1}{18}(-2 \delta^2 t^2 y -\delta^3 )+O(y^4)
\end{equation}
and the minimum eigenvalue is $\lambda=\frac{4 \delta-2yt^2}{3}$.

Clearly for positive $y$ these solutions are just approximations to the full
solution. In particular we found (\ref{Bmax}) to have the same instability than
the 2RSB free solution ($\lambda=-\frac{yt^2}{6}$).

\section{Discussions}

Let us analyse the previous results. For $p_d=0$ the two real replicas lie
on configurations belonging to different equilibrium states. As the
multiplicity of pure states
give no contributions to the entropy, the free energy per replica is
equal to that of the unconstrained system and $q(x)=q(x)^{unconstr}$ and
$p(x)=0$. IF $p_d$ increases we force the overlap between
couple of replicas to be out of
equilibrium. The free energy increases until the maximum is
reached at $p^{max}$ while if
$p_d>p^{max}$ the free energy falls until $p_d$ of order of
$q_{EA}$, where we are choosing configurations inside the
same state. Since the overwhelming majority of pairs of configurations
inside a state are at
distance $q_{EA}$ from each other, if $p_d=q_{EA}$ we have twice the energy
and twice the intra-state entropy.

At $p_d=p^{max}$ we have
$\frac{\partial F^{AB}}{\partial p_d}=\frac{dF^{AB}}{dp_d}=0$.  The
constraining force is then zero (on the S.P.) and the 2 real replicas
are unconstrained. In fact
the solution (\ref{Bmax}) is related through a replica
permutation to the solution
(\ref{sol2rsb}), because $m_1^{2RR}=m_1^{1RR}/2$.
Then $F_{max}^{2RR}=F^{1RR}_{2RSB}$.  However the constraint
still acts on the fluctuations and prevents the 2 replicas from
sliding along the unstable direction.

It is remarkable that in this way we have been able to
interpret an unstable 2RSB.
We conjecture that with more constraints one will be able to
stabilise solutions with more levels of breaking. Therefore the
linear solution, which is the limit for an infinite number of
breakings, will reappear.

The Parisi parameter $q(x)$ plays two different roles in the SK
model. On the one hand it appears in the Gibbs measure and
determines the $P_{eq}(q)$. In this case $q(x)$ must be
necessarily monotonous. On the other hand $x(q)$ also appears
in the Cavity Method where monotonicity does not seem to be
required. It remains as an intriguing open question whether in
systems as the present a non monotonous $x(q)$ could be
relevant. Perhaps solving the R real replicas by a generalisation
of the Cavity Method could elucidate this issue.

\section*{Acknowledgements}

We are grateful to S. Franz, I. Kondor, J. Kurchan and T. Temesvari for
useful discussions.

\null\newpage

\begin{figure}
\caption[]{$\Delta F$ as a function of $p_d$ plotted for $0<p_d<2p^A$,
$t=0.05$ and $y=-0.01$.}
\label{fig:1}
\end{figure}

\begin{figure}
\caption[]{$\Delta F$ as a function of $p_d$ plotted for $0<p_d<q_{EA}$,
$t=0.05$ and $y=-0.01$.}
\label{fig:2}
\end{figure}

\begin{figure}
\caption[]{$\Delta F$ as a function of $p_d$ plotted for $p^{D}$ near $q_{EA}$,
 $t=0.05$ and $y=-0.01$.}
\label{fig:3}
\end{figure}

\end{document}